# Thermal Rectification in CVD Diamond Membranes Driven by Gradient Grain Structure

Zhe Cheng[1], Brian M. Foley[1], Thomas Bougher[1], Luke Yates[1], Baratunde A. Cola[1, 2, a]

, Samuel Graham[1, 2, a]

[1]. George W. Woodruff School of Mechanical Engineering, Georgia Institute of Technology, Atlanta, Georgia 30332, USA

[2]. School of Materials Science and Engineering, Georgia Institute of Technology, Atlanta, Georgia 30332, USA

_______________________________________

[a)] Corresponding Emails: sgraham@gatech.edu; cola@gatech.edu

**ABSTRACT**

As one of the basic components of phononics, thermal diodes transmit heat current asymmetrically similar to electronic rectifiers and diodes in microelectronics. Heat can be conducted through them easily in one direction while being blocked in the other direction. In this work, we report an easily-fabricated mesoscale chemical vapor deposited (CVD) diamond thermal diode without sharp temperature change driven by the gradient grain structure of CVD diamond membranes. We build a spectral model of diamond thermal conductivity with complete phonon dispersion relation to show significant thermal rectification in CVD diamond membranes. To explain the observed thermal rectification, the temperature and thermal conductivity distribution in the CVD diamond membrane are studied. Additionally, the effects of temperature bias and diamond membrane thickness are discussed, which shed light on tuning the thermal rectification in CVD diamond membranes. The conical grain structure makes CVD diamond membranes, and potentially other CVD film structures with gradient grain structure, excellent candidates for easily-fabricated mesoscale thermal diodes without a sharp temperature change.

**KEYWORDS**: CVD diamond, thermal rectification, grain boundary, thermal diode, gradient structure

# Ⅰ. INTRODUCTION

Electron and phonon conduction are two main fundamental transport mechanisms in solid state materials, but all modern information computing processes are based on the flow of electrons in devices such as electronic transistors.[1,2] As their counterpart, phononics also has the potential to process information by manipulating heat flow like electron flow.[3] Thermal diodes are basic components of phononics, which aim to control heat current similar to electronic diodes in microelectronics.[4] For an electronic diode, its electrical resistance is small when applying a bias (electrical voltage) in one direction while the electrical resistance becomes very large when applying the bias in the other direction. This controls the electrical current to flow through the electronic diode asymmetrically. Similarly, when applying a bias (temperature difference) across a thermal diode in opposite directions, the thermal resistance is small when applying the bias in one direction while the thermal resistance becomes large in the other direction. Here, we define the thermal rectification as the ratio of the thermal resistance difference in the two directions and the smaller thermal resistance.[5,6]

In the past two decades, interest in thermal rectification has been growing rapidly because of its potential applications in information computation, thermal control, and energy conversion.[1,2,4] Both theoretical and experimental explorations have been demonstrated by graded mass density,[7] two components with different materials,[8,9] asymmetric geometry,[10-14] phase change materials,[3,15,16] thermal radiation,[17-19] holes or graded doping.[20,21] Most of these demonstrations require very complicated nanofabrication techniques or have an interface which leads to a localized temperature drop.[20] These scenarios suggest that it would be interesting to observe large thermal rectification in a one-material configuration that can be fabricated easily.

Chemical vapor deposited (CVD) diamonds have been extensively studied to dissipate heat for applications of thermal management in power electronics.[22-26] To grow CVD diamond on a substrate, the substrate needs to be mechanically abraded with diamond powders as seeds. Microwave-enhanced chemical vapor deposition method is used to grow diamond from these seeds by using a mixture of hydrogen and methane.[27] The diamond crystals grow in a columnar structure and expand laterally as the film thickness increases from the diamond-substrate interface.[28] Consequently, the diamond crystal size at the diamond-substrate interface is much smaller than that at the growth interface. This gradient grain structure results in different dominant phonon scattering mechanisms that have different temperature dependence in the two sides of the diamond membrane. When applying a temperature bias across the diamond membrane in two different directions, we can observe thermal rectification (different thermal resistances).

In contrast to the approaches of fabricated defects, holes, or asymmetric shapes, a CVD diamond membrane itself has conical grain structure and does not need extra fabrication. The gradient phonon confinement (grain boundary scattering) resulting from the gradient grain sizes makes it an excellent candidate for thermal rectification. In this work, we report a large theoretical thermal rectification in CVD diamond membranes by building a spectral thermal conductivity model based on complete phonon dispersion relation of diamond. This concept can be applied to other CVD growth materials where there is a strong change in the temperature dependence of thermal conductivity through the material.

## Ⅱ. THERMAL CONDUCTIVITY MODEL

For mesoscale heat conduction modelling, the length scale is too large for Molecular Dynamics simulation and too small for assuming diffusive conduction. Empirical formulas or modified Callaway models were usually used to describe the thermal conductivity of CVD diamond. Sood *et al.* used a gray kinetic model to model thermal conductivity by assuming an effective mean free path.[29] Several modified Callaway models were also used to model diamond thermal conductivity with assumptions of constant phonon velocity, and an ideal Debye-type phonon spectrum.[30-33] Here, we model the thermal conductivity of CVD diamond with complete phonon dispersion relation without these assumptions, similar to the method used by Mingo *et al.* for silicon nanowires and Foley, *et al.* for nano-grained SrTiO₃ thin films.[34,35] The specific heat and thermal conductivity are calculated by

$$C_v = \frac{1}{2\pi^2} \sum_j \int_k \hbar \omega_j \frac{\partial f_{BE}}{\partial T} k^2 dk \, , \tag{1}$$

$$\kappa = \frac{1}{6\pi^2} \sum_j \int_k \hbar \omega_j \frac{\partial f_{BE}}{\partial T} v_j^2 \tau_j k^2 dk \, , \tag{2}$$

where $\kappa$ is the thermal conductivity, $j$ is the phonon polarization index, $v$ is the phonon group velocity, $\tau$ is the total scattering rate, $\omega$ is the angular frequency, $\partial f_{BE}/\partial T$ is the temperature derivative of Bose-Einstein distribution function, and $k$ represents the phonon wavevector.[35] We chose (100) direction of the phonon dispersion relation of diamond to do all the calculations because this direction is highly symmetric which can be assumed as isotropic Brillouin zone.[36] The temperature dependent specific heat of diamond was calculated and agreed excellently with experimental values,[37] as shown in FIG. S1. The total scattering rate is determined by the Matthiessen's rule,[38] given as

$$\tau_j = \left[ \frac{1}{\tau_{imp}} + \frac{1}{\tau_u} + \frac{v_j}{d} \right]^{-1},$$

(3)

where, relaxation time for impurity is $\tau_{imp} = (A\omega_j^4)^{-1}$, relaxation time for Umklapp scattering is

$\tau_u = (BT\omega_j^2 e^{-C/T})^{-1}$, $d$ is the sample size. By fitting calculated thermal conductivity to previously

reported experimental values[32], we can obtain constants $A$, $B$, and $C$ as 1e$^{-48}$ s$^3$, 2.03e$^{-20}$ s/K, and

425 K. The fitting plot is shown in FIG. S2.

For CVD diamond, the grain sizes change with distance from the nucleation side $z$, as shown

in FIG. 1. We calculate the in-plane and cross-plane crystal sizes as same as Ref. [28,29] (Formulas

7 and 8 in Ref. [29]). The local scattering rate for cross-plan thermal conduction is given by

$$\tau_j(z) = \left[ \frac{1}{\tau_{imp}} + \frac{1}{\tau_u} + \frac{v_j}{d_{zeff}(z)} + \frac{v_j}{d_{reff}(z)} \right]^{-1},$$

(4)

where, effective cross-plane grain size $d_{zeff}(z) = 0.75\, d_z(z)\, t/(1\text{-}t)$, effective in-plane grain size

$d_{reff}(z) = d_r(z)(1+p)/(1\text{-}p)$. $t$ and $p$ are constants related to transmission and specularity. We take

$t$=0.56 and $p$=0.33 which are fitted from experimental data in Ref. [28]. For impurity scattering, we

assume the impurity scattering at the nucleation interface is ten times larger than the isotope scattering ($\tau_{imp} = (10A\omega_j^4)^{-1}$) and decays to $\tau_{imp} = (A\omega_j^4)^{-1}$ in 2 μm with second order polynomial

function with the distance from the nucleation interface ($z$).[20] Then the thermal conductivity of

CVD diamond is a function of distance from nucleation interface $z$ and temperature $T$.

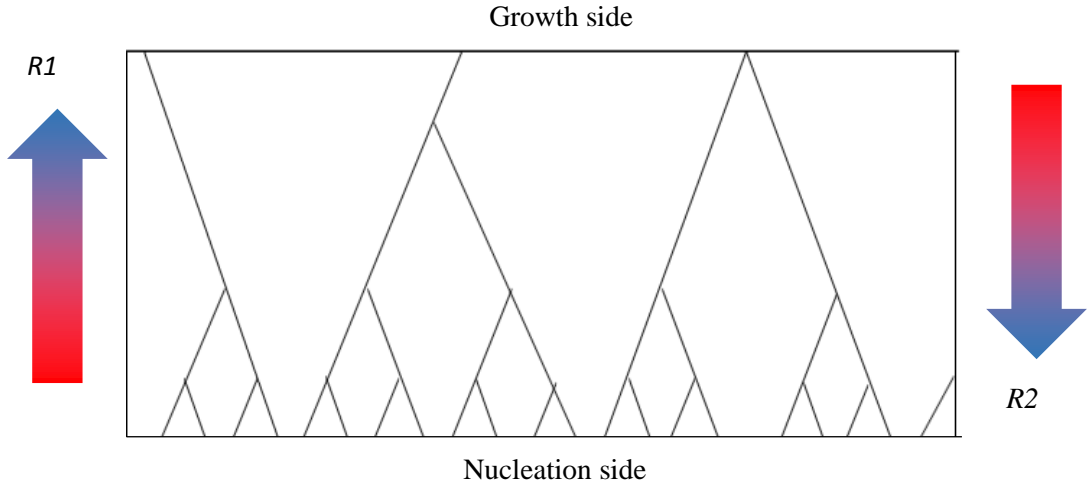

FIG. 1. Schematic diagram of grain structure of the CVD diamond membrane. *R1* and *R2* are thermal resistances of the diamond membrane when heat flux goes from the nucleation side to the growth side and from the growth side to the nucleation side.

A finite element method is easily used to obtain the temperature and thermal conductivity based on Fourier's law, as shown in FIG. 2. The diamond membrane is divided into 2000 layers. Here, the heat transfer in the diamond membrane is a one-dimensional steady-state heat conduction. To obtain the thermal resistance $R = \Delta T/q$, if we fix the temperature of one side of the membrane as $T_1$, we need to find a certain heat flux to make the temperature of another side of the membrane as $T_{2000} = T_1 + \Delta T$. We know the distance from nucleation interface of each layer once we know the membrane thickness. The temperature of all layers are initialized as $T_1$. For every layer ($i$), its thermal conductivity ($\kappa_i$) can be obtained after we know its distance from the nucleation interface ($z_i$) and its temperature ($T_i$). By guessing a value of heat flux $q$, the temperature of each layer is updated with $T_i = T_{i-1} - q \times \Delta z/\kappa_i$. Here $\Delta z$ is the thickness of each layer. Repeat updating $T_i$ and $\kappa_i$ until they are self-consistent. After that, the temperature difference between layer 1 and layer 2000 is obtained and compared with the needed value as feedback to change the value of $q$. After we find the heat flux that makes the temperature difference between layer 1 and layer 2000 as the needed value, the thermal resistance can be obtained. This procedure also works for calculating thermal resistance in the opposite direction. We only need to change the heat flux to a negative value. The thermal rectification is the ratio of the thermal resistance difference for two opposite directions and the smaller thermal resistance.

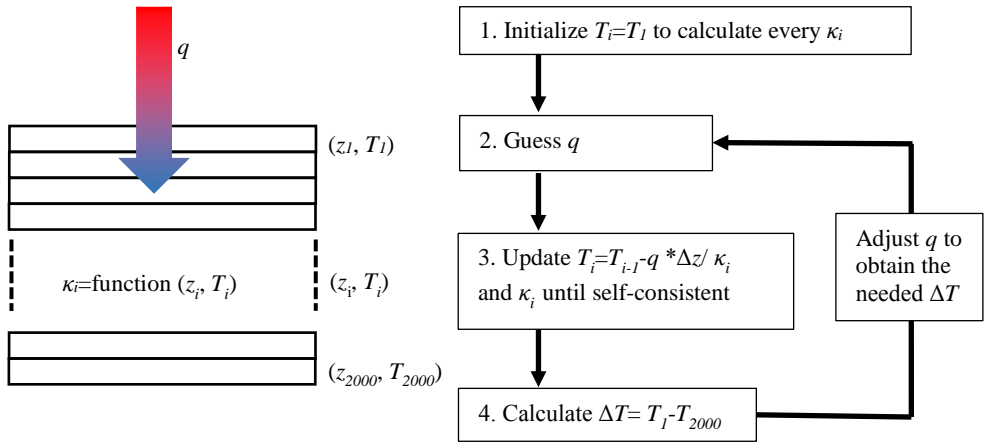

FIG. 2. Schematic diagram of the iteration process of calculating thermal rectification.

## Ⅲ. RESULTS AND DISCUSSION

FIG. 3 shows the temperature dependent local cross-plane thermal conductivity of the CVD diamond membrane with different distances from the nucleation interfaces ($z$). If taking the curve of $z$=100 μm as an example, the diamond is polycrystalline but has very large grains so phonon-phonon scatterings will dominate over grain boundary scatterings. As a result, the thermal conductivity increases with increasing temperature at low temperatures because of the sharp temperature dependence of specific heat ($T^3$). The thermal conductivity decreases with increasing temperature at high temperatures because of phonon-phonon scattering. At a given temperature, the local thermal conductivity increases with $z$ because the grain size increases with $z$. The temperature where peak thermal conductivity occurs shifts to higher temperatures with smaller $z$ because structural imperfection like grain boundary scattering and impurity scattering impedes phonon transport significantly at low temperatures. Elastic scatterings like grain boundary scatterings and impurity scatterings are temperature independent while phonon-phonon scatterings are strong temperature-dependent. So we can see strong temperature dependence of thermal conductivity for large $z$ values and weak temperature dependence for small $z$ values. This is the phenomenon we leverage here to observe thermal rectification in CVD diamond membranes. In the temperature range of 200-300 K, thermal conductivity for $z$=1 μm increases with temperature while thermal conductivity for larger $z$ values decreases with temperature. These opposite trends facilitate thermal rectification in CVD diamond membranes.

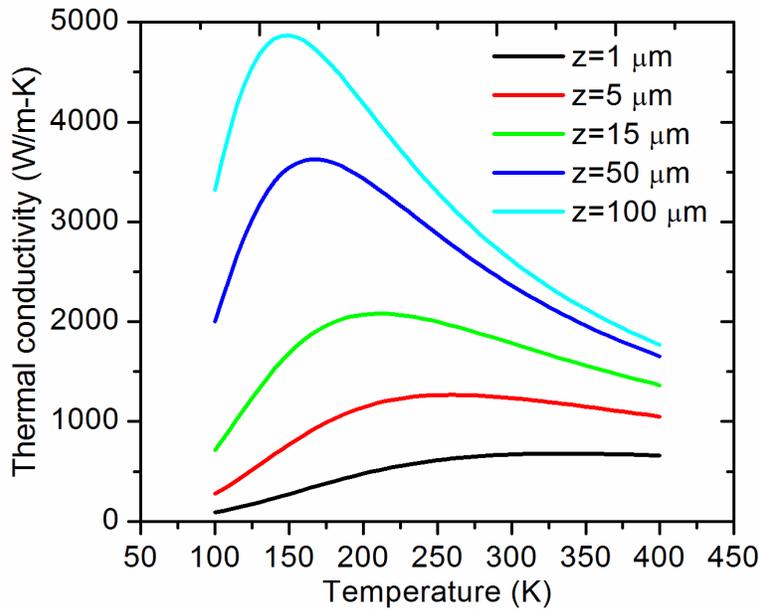

FIG. 3. Temperature dependent local cross-plane thermal conductivity of a CVD diamond membrane for different distances from nucleation side ($z$).

The value of thermal rectification is defined as $(R_2-R_1)/R_1$, as shown in FIG. 1. $R_1$ and $R_2$ are thermal resistances of the whole diamond membrane with heat flux flowing in the two opposite directions (i.e. $R_1$ from the nucleation side to the growth side, and $R_2$ from the growth side to the nucleation side). FIG. 4 shows the temperature and thermal conductivity distribution along thickness direction when applying a temperature bias of 175-375 K to a 100 μm thick diamond membrane. The slope of temperature change with $z$ is large near nucleation interface because of its low thermal conductivity, especially when the nucleation side is on the low temperature side of the bias. When the nucleation side is the cold side and the growth side is the hot side, the thermal conductivity of both the cold and hot sides are small, resulting in a large thermal resistance. When the growth side is the cold side and the nucleation side is the hot side, the thermal conductivity of both the nucleation and growth sides are large, resulting in a small thermal resistance. Only for a small part in the middle of the membrane, its thermal conductivity has the opposite trend, which decreases the thermal rectification. These differences in thermal conductivity distribution results in the preferential heat flow moving from different sides of the membrane. The total thermal resistance when heat flux flows from the nucleation side to the growth side is smaller than that when heat flux flows from the growth side to the nucleation side. As a result, the thermal rectification of this diamond membrane configuration reaches to 25%.

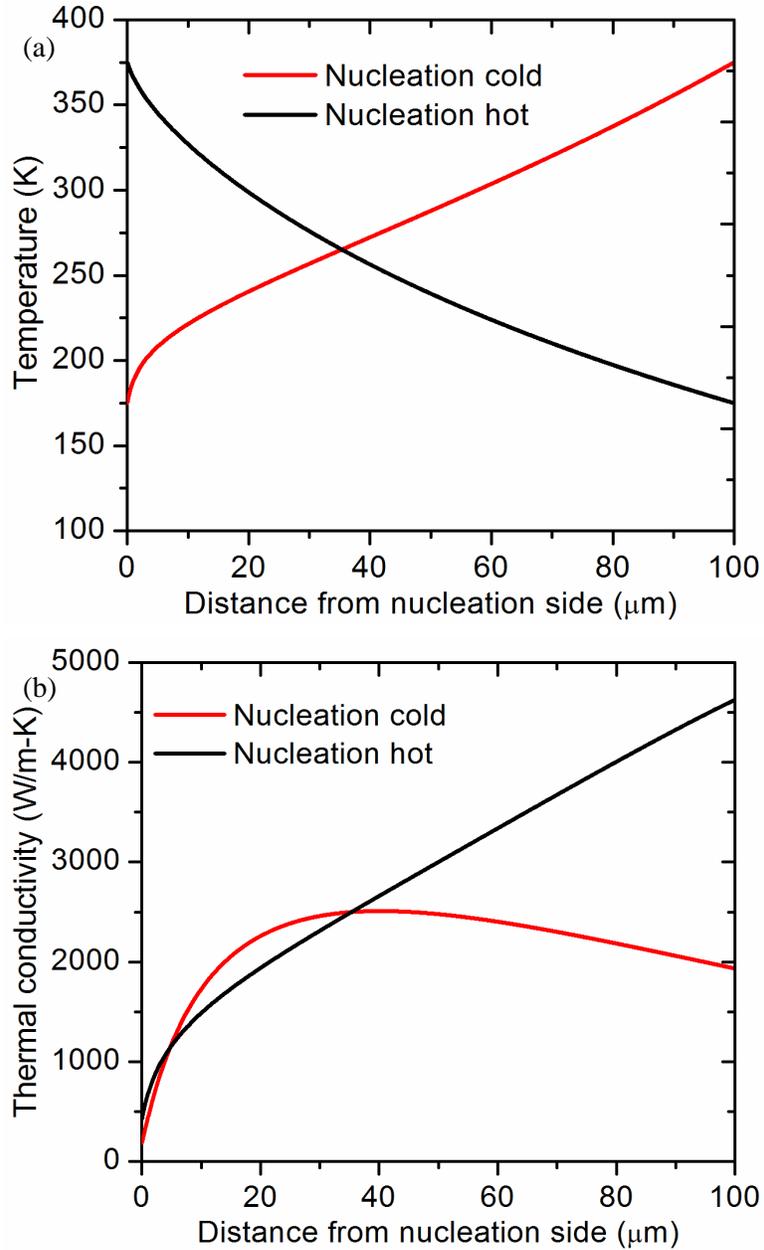

FIG. 4. Temperature (a) and thermal conductivity (b) distribution along thickness direction with a temperature bias of 175-375 K. "Nucleation cold" represents the nucleation side of the diamond membrane has a lower temperature than that at the growth side. "Nucleation hot" represents the nucleation side of the diamond membrane has a higher temperature than that at the growth side.

Membrane thickness and temperature bias affect the thermal rectification of CVD diamond membrane significantly. FIG. 5 shows thickness dependent thermal rectification with different temperature biases of 50 K, 100 K, and 200 K. The average temperature of all these three cases is 275 K. The thermal rectification increases with membrane thickness rapidly before reaching a peak. Then it decreases slowly with increasing thickness.

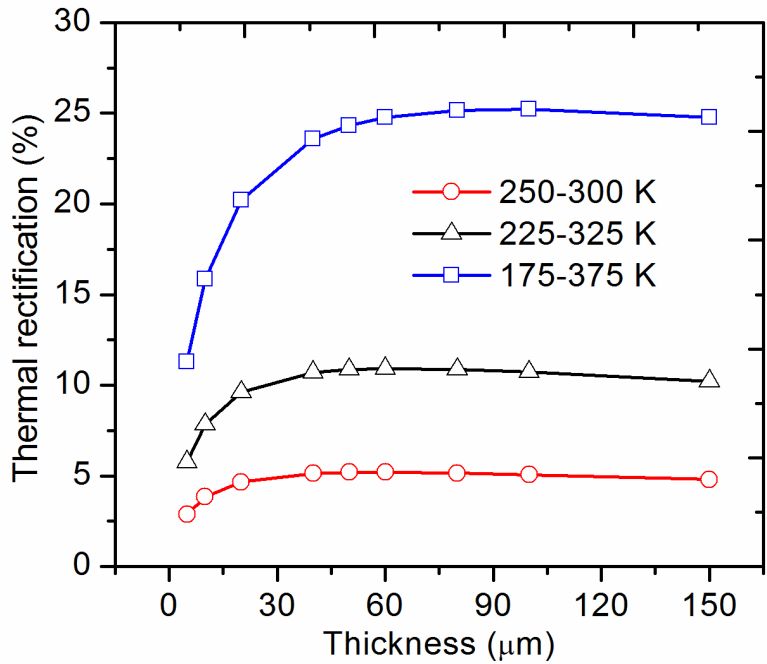

FIG. 5. Thickness dependent thermal rectification of a CVD diamond membrane with 50 K, 100 K, and 200 K temperature biases.

When the membrane is thin, grain boundary scattering plays an important role in impeding thermal transport even at growth side of the membrane. Unlike phonon-phonon scattering, structural imperfection scatterings like grain boundary scattering are elastic scatterings which are temperature independent. When grain boundary scatterings are dominant over phonon-phonon scatterings, the thermal conductivity is less temperature dependent. When the temperature bias is applied, thermal rectification becomes small because the thermal conductivity is weakly temperature dependent and does not change much with different temperature distribution. As the thickness increases, the large grain size impedes phonon transport to a diminishing degree. Phonon-phonon scattering becomes the dominant mechanism affecting thermal transport, which is highly temperature dependent. This leads to an increasing rectification ratio. But as the membrane thickness keeps increasing, the temperature gradient across the membrane thickness direction becomes increasingly small. The middle part of the membrane does not contribute to thermal rectification. This leads to a slow decrease of thermal rectification when the membrane thickness keeps increasing. Additionally, the thermal rectification increases with the applied temperature bias. The thermal rectification with temperature bias of 100 K is close to twice as that with temperature bias of 50 K. However, the thermal rectification with temperature bias of 200 K is larger than twice of that with temperature bias of 100 K. This is due to the very large temperature dependence of diamond thermal conductivity at low temperatures (175-225 K). To observe thermal rectification in the diamond membranes, high heat flux is required (several kW/mm$^2$). Additionally, to distinguish the effects of grain boundary and defects near the nucleation interface we added in the model on the thermal rectification, we calculate the thermal rectification in both cases (with and without the extra defects), as shown in FIG. 6. The two lines overlap with each other for both cases so the added defects in the model does not affect the thermal rectification. The observed thermal rectification is mainly due to gradient grain structure.

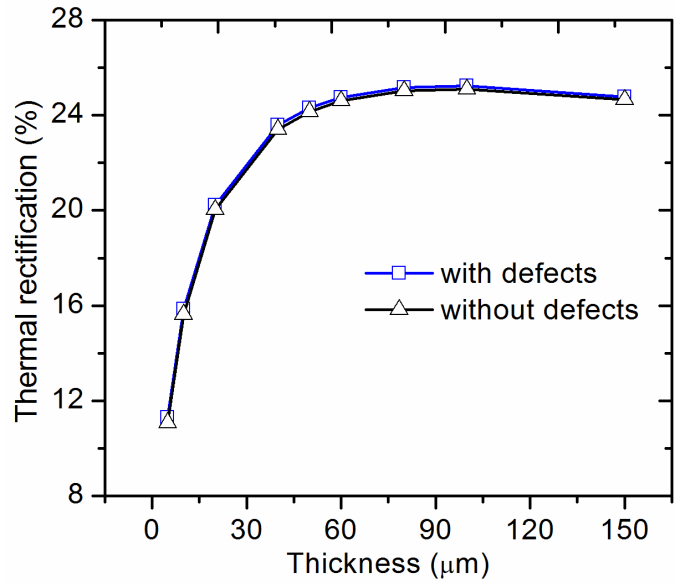

FIG. 6. Thermal rectification for diamond membranes with and without considering added defects near the nucleation interface in the model. The temperature bias is 175-375 K.

## IV. CONCLUSION

As one of the basic components of phononics, thermal diodes transmit heat current asymmetrically similar to electronic rectifiers and diodes in microelectronics. In this work, we report an easily-fabricated mesoscale chemical vapor deposited (CVD) diamond thermal diode without sharp temperature change driven by the gradient grain structure of the CVD diamond membrane. We developed a spectral thermal conductivity model of the thermal conductivity of CVD diamond with complete phonon dispersion relation to show significant thermal rectification in CVD diamond membranes. With the use of our model we find thermal rectification reaches 5%, 11%, and 25% with temperature biases of 50K, 100K, and 200K, which suggests that CVD diamond could serve as practical thermal diodes. To explain the observed thermal rectification, we studied the temperature and thermal conductivity distribution in the CVD diamond membrane. Additionally, we discussed the effects of temperature bias and diamond membrane thickness, which shed light on tuning thermal rectification in CVD diamond membranes. The conical grain structure makes CVD diamond membranes, and potentially other CVD film structures with gradient grain structure, excellent candidates for easily-fabricated mesoscale thermal diodes without a sharp temperature change.

## ACKNOWLEDGEMENTS

The authors would like to acknowledge the funding support from DARPA Thermal Transport in Diamond Thin Films for Electronic Thermal Management initiative under contract no. FA8650-15C-7517.

# REFERENCES

1       Li, N. *et al.* Colloquium: Phononics: Manipulating heat flow with electronic analogs and beyond. *Reviews of Modern Physics* **84**, 1045 (2012).

2       Wehmeyer, G., Yabuki, T., Monachon, C., Wu, J. & Dames, C. Thermal diodes, regulators, and switches: Physical mechanisms and potential applications. *Applied Physics Reviews* **4**, 041304 (2017).

3       Zhang, T. & Luo, T. Giant Thermal Rectification from Polyethylene Nanofiber Thermal Diodes. *small* **11**, 4657-4665 (2015).

4       Li, B., Wang, L. & Casati, G. Thermal diode: Rectification of heat flux. *Physical review letters* **93**, 184301 (2004).

5       Hu, S., An, M., Yang, N. & Li, B. A series circuit of thermal rectifiers: an effective way to enhance rectification ratio. *small* **13** (2017).

6       Yang, N., Zhang, G. & Li, B. Carbon nanocone: a promising thermal rectifier. *Applied Physics Letters* **93**, 243111 (2008).

7       Chang, C., Okawa, D., Majumdar, A. & Zettl, A. Solid-state thermal rectifier. *Science* **314**, 1121-1124 (2006).

8       Kobayashi, W., Teraoka, Y. & Terasaki, I. An oxide thermal rectifier. *Applied Physics Letters* **95**, 171905 (2009).

9       Hu, M., Goicochea, J. V., Michel, B. & Poulikakos, D. Thermal rectification at water/functionalized silica interfaces. *Applied Physics Letters* **95**, 151903 (2009).

10      Yang, N., Zhang, G. & Li, B. Thermal rectification in asymmetric graphene ribbons. *Applied Physics Letters* **95**, 033107 (2009).

11    Hu, J., Ruan, X. & Chen, Y. P. Thermal conductivity and thermal rectification in graphene nanoribbons: a molecular dynamics study. *Nano letters* **9**, 2730-2735 (2009).

12    Liu, Y.-Y., Zhou, W.-X. & Chen, K.-Q. Conjunction of standing wave and resonance in asymmetric nanowires: a mechanism for thermal rectification and remote energy accumulation. *Scientific reports* **5** (2015).

13    Cartoixà, X., Colombo, L. & Rurali, R. Thermal rectification by design in telescopic Si nanowires. *Nano letters* **15**, 8255-8259 (2015).

14    Ma, H. & Tian, Z. Significantly High Thermal Rectification in an Asymmetric Polymer Molecule Driven by Diffusive versus Ballistic Transport. *Nano letters* (2017).

15    Chen, R. *et al.* Controllable thermal rectification realized in binary phase change composites. *Scientific reports* **5**, 8884 (2015).

16    Zhu, J. *et al.* Temperature-gated thermal rectifier for active heat flow control. *Nano letters* **14**, 4867-4872 (2014).

17    Otey, C. R., Lau, W. T. & Fan, S. Thermal rectification through vacuum. *Physical Review Letters* **104**, 154301 (2010).

18    Wang, L. & Zhang, Z. Thermal rectification enabled by near-field radiative heat transfer between intrinsic silicon and a dissimilar material. *Nanoscale and Microscale Thermophysical Engineering* **17**, 337-348 (2013).

19    Elzouka, M. & Ndao, S. High Temperature Near-Field NanoThermoMechanical Rectification. *Scientific Reports* **7** (2017).

20    Dettori, R., Melis, C., Rurali, R. & Colombo, L. Thermal rectification in silicon by a graded distribution of defects. *Journal of Applied Physics* **119**, 215102 (2016).

21    Wang, H. *et al.* Experimental study of thermal rectification in suspended monolayer graphene. *Nature Communications* **8** (2017).

22    Anaya, J. *et al.* Simultaneous determination of the lattice thermal conductivity and grain/grain thermal resistance in polycrystalline diamond. *Acta Materialia* 139 (2017): 215-225.

23    Cheaito, Ramez, Aditya Sood, Luke Yates, Thomas L. Bougher, Zhe Cheng, Mehdi Asheghi, Samuel Graham, and Kenneth Goodson. "Thermal conductivity measurements on suspended diamond membranes using picosecond and femtosecond time-domain thermoreflectance." In Thermal and Thermomechanical Phenomena in Electronic Systems (ITherm), 2017 16th IEEE Intersociety Conference on, pp. 706-710. IEEE, 2017.

24    Rougher, Thomas L., Luke Yates, Zhe Cheng, Baratunde A. Cola, Samuel Graham, Ramez Chaeito, Aditya Sood, Mehdi Ashegi, and Kenneth E. Goodson. "Experimental considerations of CVD diamond film measurements using time domain thermoreflectance." In Thermal and Thermomechanical Phenomena in Electronic Systems (ITherm), 2017 16th IEEE Intersociety Conference on, pp. 30-38. IEEE, 2017.

25    Yates, Luke, Aditya Sood, Zhe Cheng, Thomas Bougher, Kirkland Malcolm, Jungwan Cho, Mehdi Asheghi et al. "Characterization of the Thermal Conductivity of CVD Diamond for GaN-on-Diamond Devices." In Compound Semiconductor Integrated Circuit Symposium (CSICS), 2016 IEEE, pp. 1-4. IEEE, 2016.

26    Yates, Luke, Ramez Cheaito, Aditya Sood, Zhe Cheng, Thomas Bougher, Mehdi Asheghi, Kenneth Goodson et al. "Investigation of the Heterogeneous Thermal Conductivity in Bulk CVD Diamond for Use in Electronics Thermal Management." In ASME 2017 International Technical Conference and Exhibition on Packaging and Integration of Electronic and Photonic Microsystems collocated with the ASME 2017 Conference on Information Storage and Processing Systems, pp. V001T04A014-V001T04A014. American Society of Mechanical Engineers, 2017.

27      Silva, F., Hassouni, K., Bonnin, X. & Gicquel, A. Microwave engineering of plasma-assisted CVD reactors for diamond deposition. *Journal of physics: condensed matter* **21**, 364202 (2009).

28      Cheng, Zhe, Thomas Bougher, Tingyu Bai, Steven Y. Wang, Chao Li, Luke Yates, Brian M. Foley, Mark Goorsky, Baratunde A. Cola, and Samuel Graham. "Probing Growth-Induced Anisotropic Thermal Transport in CVD Diamond Membranes by Multi-frequency and Multi-spot-size Time-Domain Thermoreflectance." arXiv preprint arXiv:1708.00405 (2017).

29      Sood, A. *et al.* Anisotropic and inhomogeneous thermal conduction in suspended thin-film polycrystalline diamond. *Journal of Applied Physics* **119**, 175103 (2016).

30      Berman, R. Thermal conductivity of isotopically enriched diamonds. *Physical Review B* **45**, 5726 (1992).

31      Onn, D., Witek, A., Qiu, Y., Anthony, T. & Banholzer, W. Some aspects of the thermal conductivity of isotopically enriched diamond single crystals. *Physical review letters* **68**, 2806 (1992).

32      Wei, L., Kuo, P., Thomas, R., Anthony, T. & Banholzer, W. Thermal conductivity of isotopically modified single crystal diamond. *Physical Review Letters* **70**, 3764 (1993).

33      Olson, J. *et al.* Thermal conductivity of diamond between 170 and 1200 K and the isotope effect. *Physical Review B* **47**, 14850 (1993).

34    Mingo, N. Calculation of Si nanowire thermal conductivity using complete phonon dispersion relations. *Physical Review B* **68**, 113308 (2003).

35    Foley, B. M. *et al.* Thermal conductivity of nano-grained SrTiO 3 thin films. *Applied Physics Letters* **101**, 231908 (2012).

36    Tubino, R., Piseri, L. & Zerbi, G. Lattice dynamics and spectroscopic properties by a valence force potential of diamondlike crystals: C, Si, Ge, and Sn. *The Journal of Chemical Physics* **56**, 1022-1039 (1972).

37    Raman, C. The heat capacity of diamond between 0 and 1000° K. *Proceedings Mathematical Sciences* **46**, 323-332 (1957).

38    Kittel, C. *Introduction to solid state physics*. (Wiley, 2005).